# On the Complexities of Testing for Compliance with Human Oversight Requirements in AI Regulation

Markus Langer [1[0000-0002-8165-1803]] and Veronika Lazar [2] and Kevin Baum [3[0000-0002-6893-573X]]

[1] University of Freiburg, Department of Psychology, Freiburg, Germany

[2] German Federal Office for Information Security, Saarbrücken, Germany

[3] German Research Center for Artificial Intelligence, Responsible AI and Machine Ethics Research Group, Saarbrücken, Germany

**Abstract.** Human oversight requirements are a core component of the European AI Act and in AI governance. In this paper, we highlight key challenges in testing for compliance with these requirements. A central difficulty lies in balancing simple, but potentially ineffective checklist-based approaches with resource-intensive and context-sensitive empirical testing of the effectiveness of human oversight of AI. Questions regarding when to update compliance testing, the context-dependent nature of human oversight requirements, and difficult-to-operationalize standards further complicate compliance testing. We argue that these challenges illustrate broader challenges in the future of sociotechnical AI governance, i.e. a future that shifts from ensuring "good" technological products to "good" sociotechnical systems.

## 1 Introduction

Testing for compliance with emerging legislation regarding Artificial Intelligence (AI) such as the European AI Act will be a major task for providers and deployers of AI-based systems when these systems are used for high-risk tasks [1], [2]. Some aspects of this compliance testing will resemble traditional auditing processes for classical software systems and other technologies governed by product safety regulations [1]. For instance, verifying whether AI systems have adequate documentation, ensuring cybersecurity, testing for data protection, and evaluating the accuracy of system outputs could all be achieved by defining standards and quality thresholds. Eventually, compliance testing may also draw on established best practices, such as checklist-based approaches to assess whether implemented processes and technologies comply with standards set by norming bodies. Of course, new testing procedures will also be required to assess robustness and fairness according to various criteria, particularly with regard to the immediate safety of (semi-)autonomous, agentic systems. New testing procedures and benchmarks will emerge, and services and infrastructure will build around them.

However, the AI Act and other emerging AI regulations introduce requirements that go beyond the technical characteristics of the system and requirements that traditional compliance testing methods cannot easily address. One such key requirement is effective human oversight, as outlined in Article 14 of the AI Act [3], [4]. Various countries, including Argentina, Bahrain, Uganda, and South Africa, already enforce similar but less specific requirements for human involvement in AI-driven decision-making [3]. Effective human oversight in the sense of the AI Act includes specific sub-requirements, such as ensuring that human oversight personnel remain aware of their tendency to over-rely on outputs produced by a high-risk AI system (e.g., automation bias; [5]) and that they properly understand the relevant capacities and limitations of the high-risk AI system to adequately monitor its operation (see Article 14 AI Act).

In this paper we claim that a key challenge of testing for compliance with human oversight requirements lies in balancing simple but potentially ineffective checklist-based approaches with resource-intensive and context-sensitive empirical testing of the effectiveness of human oversight of AI. Questions regarding when to update compliance testing, the context-dependent nature of human oversight requirements, and difficult-to-operationalize standards for when oversight is truly "effective" further complicate compliance testing [6]. The intricate sociotechnical interplay of technical aspects, individual factors, and environmental conditions that determine human oversight effectiveness adds additional complexity [8]. In fact, research suggests that human oversight requirements are particularly difficult to operationalize and test (see e.g., [7]). We argue that all of this illustrates broader challenges in the future of sociotechnical AI governance, i.e. a future that shifts from ensuring "good" technological products (e.g., safe products) to "good" sociotechnical systems (e.g., safe human-AI interactions in a specific context).

## 2 The Possible Future of Testing for Compliance with Human Oversight Requirements

The European AI Act establishes requirements for AI systems classified as "high-risk," including those used in (critical areas of) education, public administration, hiring, credit scoring, and medicine. One such requirement is human oversight, as specified in Article 14 (see Appendix A for the full text). It states that "human oversight shall aim to prevent or minimise the risks to health, safety or fundamental rights that may emerge when a high-risk AI system is used in accordance with its intended purpose or under conditions of reasonably foreseeable misuse, in particular where such risks persist despite the application of other requirements set out in this Section." These other requirements, detailed in Articles 9-13 and Article 15 of the AI Act, cover risk management, data governance, technical documentation, record keeping, transparency, accuracy, robustness, and cybersecurity.

Some key requirements of Article 14 include that human oversight personnel should be able to understand the capacities and limitations of AI systems and correctly interpret outputs. They should remain aware of their tendency for automation bias (which, according to the AI Act, refers to overtrust in AI outputs; but see [8] for the complexities and dynamics of concepts associated with trust), decide when not to use AI outputs, and override decisions when necessary. They are also supposed to intervene or interrupt a system, for example, using a stop button or a similar mechanism to halt the system in a safe state.

Standards and norms are currently being developed to guide compliance testing for human oversight requirements, including the Trustworthiness Framework developed by CEN/CENELEC and the ISO/IEC CD 42105. Part of these standards and norms will be informed by Article 14 of the AI Act and broader international governance perspectives on human oversight. While still in development, we anticipate a future of compliance testing for human oversight along a continuum between simple, checklist-based approaches and empirical testing of the effectiveness of human oversight in specific contexts.

A checklist approach would follow the model of existing compliance testing methods [1], [7] translating Article 14 requirements into assessable items for internal or external auditors. Checklist items inspired by Article 14 might include: "The human oversight person has been made aware of their tendency to overtrust outputs of AI-based systems", "The human oversight person has received adequate training that enables them to understand the capacities and limitations of the AI-based system they oversee", or "There is a stop button that allows the human oversight person to intervene in the operation of the AI-based system". Clearly, this list is not exhaustive and the requirements would need to be refined. While such checklists could provide a straightforward compliance mechanism, they may fall short of the AI Act's broader goal of *effectively* mitigating AI-related risks [1]. Moreover, developing an exhaustive checklist will be challenging. The examples above are direct translations from Article 14. Clearly there can be an infinite number of requirements with varying degrees of specificity, for instance, requirements concerning the person who will be the human oversight person (e.g., specific skills they must possess), or work design of human oversight jobs (e.g., specific maximum durations for human oversight tasks) [9].

Empirical testing of the effectiveness of human oversight in specific contexts could address some of the limitations of checklists. This approach would require testing the actual effectiveness of human oversight in high-risk contexts and empirically demonstrating compliance with AI regulatory requirements [3]. It could involve studies where human oversight personnel monitor AI systems for a set duration, assessing whether they detect erroneous or problematic outputs, intervene in system operation when necessary, and accurately override inadequate AI-generated decisions. Another option could involve comparing different human oversight designs to determine which best meets regulatory requirements. Human oversight design, as outlined by Sterz et al. [9] is a sociotechnical design question. It encompasses technical aspects (e.g., optimizing explainability approaches to support and amplify human oversight), individual factors (e.g., selecting and training oversight personnel), and environmental conditions (e.g., job design and working conditions). For instance, a controlled experiment could test various explainability approaches to assess which most effectively supports oversight [10], [11], [12]. The main advantage of this approach is that it provides empirical evidence on the effectiveness of human oversight and how to optimize it.

However, this empirical approach demands significant resources for planning, conducting, analyzing, and interpreting studies. Empirical testing for the effectiveness of human oversight requires expertise with empirical methods and user studies. Researchers and practitioners, for instance, with a background in human-computer interaction or psychology will be required to adequately conduct empirical testing, interpreting the results, and providing recommendations on how to optimize human oversight design. Moreover, transferring insights across contexts may be challenging, as oversight effectiveness can be – and typically is expected to be – highly context-sensitive. In general, empirical testing scales poorly because it requires effort to conduct empirical testing with actual human oversight personnel, and because it likely cannot be shifted from the token or deployment level to the type or provider licensing level due to the context-sensitivity of effective human oversight and the resulting loss of transferability of insights. Furthermore, deriving reliable conclusions often requires multiple studies (e.g., on the effects of different explainability approaches), suggesting that oversight requirements may need to be informed by high-quality meta-analyses that synthesize findings across studies for broader applicability.

Checklist-based approaches and empirical approaches are clearly not the only possible options for testing compliance with human oversight requirements, but they illustrate a fundamental tension in AI governance. Traditional checklist approaches offer efficiency and standardization but risk treating oversight as a technological feature to be verified rather than a sociotechnical capability to be validated. They may produce inconclusive results, such as when oversight mechanisms are formally in place but key performance indicators conflict, when standard checklist items cannot capture context-specific adequacy, or when there are gaps between documented policies and actual practice. Empirical approaches, by contrast, can validate the actual effectiveness of human oversight but are resource-intensive and context-sensitive and thus difficult to standardize.

This tension reflects the broader challenge of shifting from evaluating "good" technological products against standardized criteria to validating "good" sociotechnical systems where effectiveness emerges from dynamic interactions between humans, technology, and specific contexts. Hybrid approaches may offer a way forward: when checklist-based testing produces inconclusive results or reveals gaps between formal compliance and effective oversight, empirical validation could help bridge the measurement-reality divide. Clearly, such hybrid approaches raise their own questions about standardization as well as practical challenges such as resource allocation regarding compliance testing.

Furthermore, any form of checklist-based and empirical approaches faces additional challenges. First, it remains unclear when and how frequently human oversight processes need to be reevaluated. Should compliance assessments be conducted regularly or should they be triggered by evidence of non-compliance? Is it required to reevaluate compliance after each AI system update or after personnel changes?

Second, not only is the effectiveness of human oversight context-sensitive, the human oversight requirements themselves may also be context-dependent. For instance, oversight requirements may vary depending on the risk associated with the application context. Stricter oversight requirements may apply to AI used in the public sector compared to the private sector. Additionally, the required skills, expertise, and tasks of human oversight personnel can also vary significantly [7]. In real-time contexts, such as for the oversight of autonomous vehicles, sustained vigilance over long periods may be necessary, whereas in areas like hiring and credit scoring, human oversight personnel may require training in ethical and moral reasoning.

Third, certain regulatory requirements may be difficult to translate into testable standards for two key reasons. First, some requirements involve psychological factors that are challenging to operationalize and assess. This includes assessing whether human oversight personnel sufficiently understand AI limitations or testing for possible automation bias of human oversight personnel [5], [6]. Second, AI regulations typically include requirements where there is disagreement about guiding definitions and no clear ground truth available. For instance, in the case of discrimination and fairness, no commonly agreed standard exists for determining when a decision is discriminatory or unfair [4], [13]. This normative uncertainty, combined with the challenges of measuring psychological factors, helps explain why compliance testing particularly struggles with requirements related to transparency, explainability, and fairness [7].

These challenges point to the even more fundamental challenge that it remains unclear what constitutes sufficiently effective human oversight. Figure 1 illustrates this challenge. One fundamental expectation is that human oversight personnel add something beneficial to AI system operations. In other words, and in line with research and AI governance efforts using this term heavily [14], adding human oversight should increase the *trustworthiness* of AI operations. While trustworthy AI has many dimensions [8], [14], [15], one key goal – explicitly mentioned in the AI Act – is to increase safety. The effectiveness of human oversight depends on characteristics of oversight personnel (e.g., skills, training), the technology (e.g., AI transparency), and the operational context (e.g., roles, tasks, organizational factors) [9]. While it seems obvious that one key goal is that human oversight should make AI operations safer and more secure than autonomous operation of AI, determining when oversight is sufficiently safe and secure remains an open question.

Without concrete and testable standards for effective human oversight, normative uncertainty will persist for providers and deployers of AI regarding their legal compliance. This issue may be particularly pronounced for small businesses that lack the resources to establish an AI compliance department [7]. This uncertainty could lead to situations where the implementation of AI-based systems will be hampered in such businesses. Moreover, without concrete standards, virtually any implementation could be considered compliant [16]. This issue is especially problematic when audits rely on post-hoc rationalizations of human oversight implementations. In hindsight, any approach to human oversight could be justified as the "sufficiently good" or even “best possible” option.

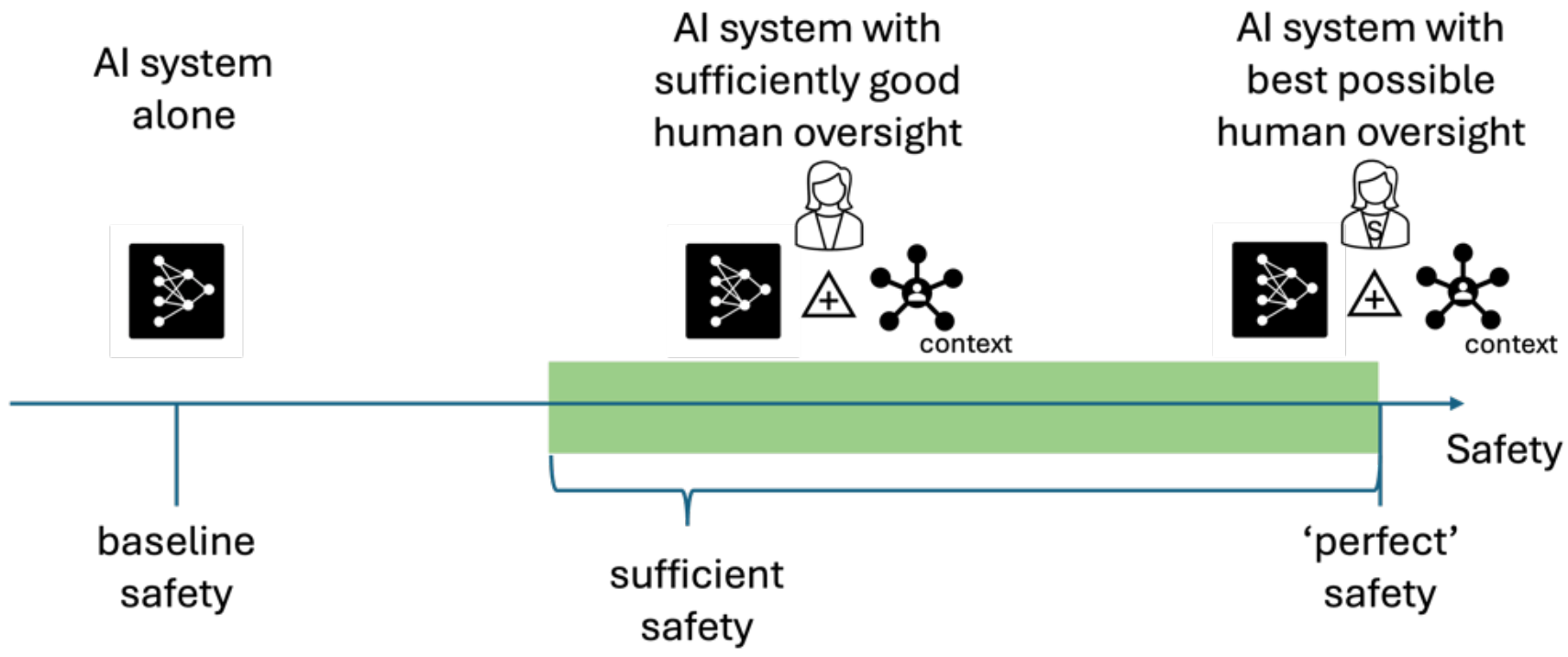

**Fig. 1.** A plausible criterion for effective human oversight is that human oversight of AI achieves a higher level of safety compared to the AI system operating alone (*baseline safety*). However, assessing safety has several layers: first, safety is inherently multi-dimensional, involving multiple, sometimes interdependent risks (e.g., an AI system may exhibit different forms of unfairness that cannot all be simultaneously eliminated). Second, even within a single safety dimension, it remains unclear what "perfect" safety would entail, making it particularly difficult to define or measure what counts as "sufficient" safety. This figure additionally highlights that safety (much like all dimensions of the effectiveness of human oversight; [9]) depends on characteristics of oversight personnel (e.g., skills, training), the technology (e.g., AI transparency), and the operational context (e.g., roles, tasks, organizational factors).

## 3 Concluding Thoughts and Next Steps

The challenges of testing for compliance with human oversight requirements reflect broader difficulties in sociotechnical AI governance. As AI governance shifts from ensuring “good” technological products to “good” sociotechnical systems, defining standards will be complex, particularly when psychological concepts are involved. Beyond human oversight requirements, another key example relates to emotion recognition systems [17], [18]. According to the AI Act, AI systems that automatically infer emotions (e.g., sadness) in high-risk contexts are prohibited but inferring physical states (e.g., fatigue) is permitted. This raises questions such as: Is fatigue a purely physical state from lack of sleep or a symptom of depression? If linked to depression, would its detection be permitted? These questions seem relevant to governing respective AI products but they can only be adequately addressed by taking a sociotechnical perspective that addresses the intricate interplay between technical design (e.g., what model is used to infer emotions?), human factors (e.g., how do verbal, nonverbal, and paraverbal behavior relate to emotions?), and contextual considerations (e.g., what inferences about emotions are adequate at different workplaces?).

As outlined before, additional difficulties arise when AI governance seeks to mitigate risks for which no clear ground truth exists (e.g., risks of discrimination [7]) or when it remains uncertain whether risks have been effectively mitigated. For instance, was a fairness monitoring tool truly successful if it detects only one specific type of fairness violation in AI outputs [4], [19], [20]? Again, this emphasizes the need for a sociotechnical perspective that addresses technical design (e.g., for what kind of fairness metric to calibrate AI-based decisions), human factors (e.g., training in the detection of fairness issues), and contextual considerations (e.g., what kind of fairness definition is appropriate in a given context)?

The next steps in testing for compliance with AI regulation are currently being developed. Standardization and norming bodies are working to operationalize the requirements outlined in AI regulation. We anticipate that emerging standards and norms will fuel the debate on how to effectively test for compliance. The challenges outlined in this article will continue to require input from researchers, practitioners, and policymakers to ensure that AI governance effectively reduces the risks associated with AI systems while enhancing safety in their implementation without placing an undue burden on providers and deployers through resource-intensive, context-dependent compliance testing.

In the case of human oversight, key tasks for the near future are to (a) establish a middle ground or a feasible combination between checklists and empirical testing, (b) develop standards and norms that are informed by and adapt to the latest research in human-computer interaction, psychology, and related fields on effective human oversight [21], such as methods for preventing automation bias or effectively preparing and supporting humans to detect inaccurate and problematic outputs, and (c) evaluate the impact of human oversight requirements in AI practice. Finally, we want to highlight the crucial importance of expertise on the human factor in human-AI interaction for designing and testing for the effectiveness of human oversight. As AI governance evolves beyond technological improvement to optimizing sociotechnical systems for high-risk tasks, we believe research(ers) from psychology, human-computer interaction and related fields should play a key role in providing insights on how to optimize the technology, how to design the jobs and environments where humans and AI-based systems interact, and how to prepare and support human oversight personnel. This ensures that expertise and perspectives on the human factors in AI augments the expertise and perspectives that already play a key role in AI governance such as legal sciences, ethics, machine learning and other technical and engineering perspectives. This would then also substantiate commonly phrased claims (particularly in Europe) that AI governance aims for "human-centered AI" implementation that in practice often lacks to consult and integrate expertise on the "human" factors.

**Acknowledgments.** This work was partially funded by DFG grant 389792660 as part of TRR 248 CPEC – Center for Perspicuous Computing (Kevin Baum & Markus Langer), by the project TITAN – Technologische Intelligenz zur Transformation, Automatisierung und Nutzerorientierung des Justizsystems funded by the Daimler and Benz Foundation (grant no. 45-06/24) (Markus Langer), by the European Regional Development Fund (ERDF) and the Saarland within the scope of the project (To)CERTAIN – Towards a Center for European Research in Trusted AI, and received support from the German Federal Ministry of Education and Research (BMBF) as part of the project MAC-MERLin (Grant Agreement No. 01IW24007) (Kevin Baum).

**Disclosure of Interests.** The authors have no competing interests to declare that are relevant to the content of this article.

## References

[1] K. Lam, B. Lange, B. Blili-Hamelin, J. Davidovic, S. Brown, and A. Hasan, "A framework for assurance audits of algorithmic systems," in *The 2024 ACM Conference on Fairness, Accountability, and Transparency*, Rio de Janeiro Brazil: ACM, Jun. 2024, pp. 1078–1092. doi: 10.1145/3630106.3658957.

[2] J. Mökander, M. Axente, F. Casolari, and L. Floridi, "Conformity assessments and post-market monitoring: A guide to the role of auditing in the proposed European AI regulation," *Minds & Machines*, vol. 32, no. 2, pp. 241–268, Jun. 2022, doi: 10.1007/s11023-021-09577-4.

[3] B. Green, "The flaws of policies requiring human oversight of government algorithms," *Computer Law & Security Review*, vol. 45, p. Article 105681, Jul. 2022, doi: 10.1016/j.clsr.2022.105681.

[4] M. Langer, K. Baum, and N. Schlicker, "Effective human oversight of AI-Based systems: A signal detection perspective on the detection of inaccurate and unfair outputs," *Minds and Machines*, 2024, doi: 10.1007/s11023-024-09701-0.

[5] H. Ruschemeier and L. Hondrich, "Automation bias in public administration - an interdisciplinary perspective from law and psychology," *SSRN Journal*, 2024, doi: 10.2139/ssrn.4736646.

[6] J. Laux, “Institutionalised distrust and human oversight of artificial intelligence: towards a democratic design of AI governance under the European Union AI Act,” *AI & Soc*, vol. 39, no. 6, pp. 2853–2866, Dec. 2024, doi: 10.1007/s00146-023-01777-z.

[7] T. Scantamburlo *et al.*, “Software systems compliance with the AI Act: Lessons learned from an international challenge,” in *Proceedings of the 2nd International Workshop on Responsible AI Engineering*, Lisbon Portugal: ACM, Apr. 2024, pp. 44–51. doi: 10.1145/3643691.3648589.

[8] N. Schlicker, A. Uhde, K. Baum, M. C. Hirsch, and M. Langer, “How do we assess the trustworthiness of AI: Introducing the Trustworthiness Assessment Model (TrAM),” *Computers in Human Behavior*, 2025, doi: 10.31234/osf.io/qhwvx.

[9] S. Sterz *et al.*, “On the quest for effectiveness in human oversight: Interdisciplinary perspectives,” in *The 2024 ACM Conference on Fairness, Accountability, and Transparency*, Rio de Janeiro Brazil: ACM, Jun. 2024, pp. 2495–2507. doi: 10.1145/3630106.3659051.

[10] A. B. Arrieta *et al.*, “Explainable artificial intelligence (XAI): Concepts, taxonomies, opportunities and challenges toward responsible AI,” *Information Fusion*, vol. 58, pp. 82–115, Jun. 2020, doi: 10.1016/j.inffus.2019.12.012.

[11] M. Langer *et al.*, “What do we want from Explainable artificial intelligence (XAI)? A stakeholder perspective on XAI and a conceptual model guiding interdisciplinary XAI research.,” *Artificial Intelligence*, vol. 296, p. 103473, 2021, doi: 10.1016/j.artint.2021.103473.

[12] V. Lai, C. Chen, A. Smith-Renner, Q. V. Liao, and C. Tan, “Towards a science of human-AI decision making: An overview of design space in empirical human-subject studies,” in *2023 ACM Conference on Fairness, Accountability, and Transparency*, Chicago IL USA: ACM, Jun. 2023, pp. 1369–1385. doi: 10.1145/3593013.3594087.

[13] N. Mehrabi, F. Morstatter, N. Saxena, K. Lerman, and A. Galstyan, “A survey on bias and fairness in machine learning,” *ACM Comput. Surv.*, vol. 54, no. 6, p. Article 115, Jul. 2021, doi: 10.1145/3457607.

[14] J. Laux, S. Wachter, and B. Mittelstadt, “Trustworthy artificial intelligence and the European Union AI act: On the conflation of trustworthiness and acceptability of risk,” *Regulation & Governance*, vol. 18, no. 1, pp. 3–32, Jan. 2024, doi: 10.1111/rego.12512.

[15] High Level Expert Group on Artificial Intelligence, *Ethics guidelines for trustworthy AI*. Brussels, Belgium: European Commission, 2019. [Online]. Available: https://ec.europa.eu/newsroom/dae/document.cfm?doc_id=60419

[16] J. Cobbe, M. S. A. Lee, and J. Singh, “Reviewable automated decision-making: A framework for accountable algorithmic systems,” in *Proceedings of the 2021 ACM Conference on Fairness, Accountability, and Transparency*, Virtual Event Canada: ACM, Mar. 2021, pp. 598–609. doi: 10.1145/3442188.3445921.

[17] A. Prégent, “Is there not an obvious loophole in the AI act’s ban on emotion recognition technologies?,” *AI & Soc*, pp. s00146-025-02289–8, Mar. 2025, doi: 10.1007/s00146-025-02289-8.

[18] P. Meinel and A. Lauber-Rönsberg, “‘How do you feel?’ – Emotionserkennung nach der KI-VO,” *Datenschutz Datensich*, vol. 49, no. 4, pp. 236–240, Apr. 2025, doi: 10.1007/s11623-025-2078-3.

[19] S. Biewer *et al.*, “Software doping analysis for human oversight,” *Formal Methods in System Design*, 2024, doi: 10.48550/arXiv.2308.06186.

[20] K. Baum *et al.*, “Taming the AI monster: Monitoring of individual fairness for effective human oversight,” in *Model Checking Software*, vol. 14624, T. Neele and A. Wijs, Eds., in Lecture Notes in Computer Science, vol. 14624. , Cham: Springer Nature Switzerland, 2025, pp. 3–25. doi: 10.1007/978-3-031-66149-5_1.

[21] J. Laux and H. Ruschemeier, “Automation bias in the AI Act: On the legal implications of attempting to de-bias human oversight of AI,” 2025, *arXiv*. doi: 10.48550/ARXIV.2502.10036.

## Appendix A

### Content of Article 14 of the European AI Act: Human Oversight

(1) High-risk AI systems shall be designed and developed in such a way, including with appropriate human-machine interface tools, that they can be effectively overseen by natural persons during the period in which they are in use.

(2) Human oversight shall aim to prevent or minimise the risks to health, safety or fundamental rights that may emerge when a high-risk AI system is used in accordance with its intended purpose or under conditions of reasonably foreseeable misuse, in particular where such risks persist despite the application of other requirements set out in this Section.

(3) The oversight measures shall be commensurate with the risks, level of autonomy and context of use of the high-risk AI system, and shall be ensured through either one or both of the following types of measures:

   (a) measures identified and built, when technically feasible, into the high-risk AI system by the provider before it is placed on the market or put into service;

   (b) measures identified by the provider before placing the high-risk AI system on the market or putting it into service and that are appropriate to be implemented by the deployer.

(4) For the purpose of implementing paragraphs 1, 2 and 3, the high-risk AI system shall be provided to the deployer in such a way that natural persons to whom human oversight is assigned are enabled, as appropriate and proportionate:

   (a) to properly understand the relevant capacities and limitations of the high-risk AI system and be able to duly monitor its operation, including in view of detecting and addressing anomalies, dysfunctions and unexpected performance;

(b) to remain aware of the possible tendency of automatically relying or over-relying on the output produced by a high-risk AI system (automation bias), in particular for high-risk AI systems used to provide information or recommendations for decisions to be taken by natural persons;

(c) to correctly interpret the high-risk AI system's output, taking into account, for example, the interpretation tools and methods available;

(d) to decide, in any particular situation, not to use the high-risk AI system or to otherwise disregard, override or reverse the output of the high-risk AI system;

(e) to intervene in the operation of the high-risk AI system or interrupt the system through a 'stop' button or a similar procedure that allows the system to come to a halt in a safe state.

(5) For high-risk AI systems referred to in point 1(a) of Annex III, the measures referred to in paragraph 3 of this Article shall be such as to ensure that, in addition, no action or decision is taken by the deployer on the basis of the identification resulting from the system unless that identification has been separately verified and confirmed by at least two natural persons with the necessary competence, training and authority.

The requirement for a separate verification by at least two natural persons shall not apply to high-risk AI systems used for the purposes of law enforcement, migration, border control or asylum, where Union or national law considers the application of this requirement to be disproportionate.